# On-Surface Hydrogen-Induced Covalent Coupling of Polycyclic Aromatic Hydrocarbons via a Superhydrogenated Intermediate

Carlos Sánchez-Sánchez[1*], José Ignacio Martínez[1], Nerea Ruiz del Arbol[1], Pascal Ruffieux[2], Roman Fasel[2,3], María Francisca López[1], Pedro de Andrés[1], José Angel Martín-Gago[1*]

[1]Materials Science Factory, Institute of Material Science of Madrid (ICMM-CSIC). 3 Sor Juana Inés de la Cruz St., 28049 Madrid (Spain).

[2]Swiss Federal Laboratories for Materials Science and Technology (Empa). Ueberlandstrasse 129, 8600 Duebendorf (Switzerland).

[3]Department of Chemistry and Biochemistry, University of Bern. Freiestrasse 3, 3012 Bern (Switzerland).

**Abstract**

The activation and subsequent covalent coupling of polycyclic aromatic hydrocarbons (PAHs) are of great interest in fields like chemistry, energy, biology, or health, among others. However, this is not a trivial process. So far, it is based on the use of catalysts that drive and increase the efficiency of the reaction. Here, we report on an unprecedented method in which the dehydrogenation and covalent coupling is thermally activated in the presence of atomic hydrogen and a surface. This mechanism, which requires of the superhydrogenation of the PAHs, has been characterized by high-resolution scanning tunnelling microscopy (STM) and rationalized by density functional theory (DFT) calculations. This work opens a door toward the formation of covalent, PAH-based, macromolecular nanostructures on low-reactive surfaces, thus facilitating its applicability.

Polycyclic aromatic hydrocarbons (PAHs) –organic compounds constituted only by carbon aromatic rings and hydrogen– are ubiquitous in society.[Blumer1976] They are routinely used in a wide variety of scientific and technological fields such as fundamental chemistry, drugs production, or petrochemical industry, among many others. Moreover, PAHs have been detected in the Universe [Tielens2008] and it has been proposed that they could have been involved in the origin of life through dehydrogenation and coupling processes.[Ehrenfreund2006] They are commonly found in air and soil as a consequence of natural processes and human activity, where they constitute an important health risk.[Haritash2009]

Dehydrogenation of PAHs, i.e. the cleavage of a C-H bond, is a common intermediate step in the formation of more complex molecules. It has also been recently proposed that chemical reactions involving dehydrogenation of superhydrogenated PAHs in space may play an important role in the formation of interstellar molecular hydrogen.[Habart2004] [Montillaud2013] Furthermore, although large polycyclic structures that may result from the coupling of smaller PAHs have been detected in space, their formation mechanism is still under intense debate. [Merino2014] [Montillaud2013]

Despite being of unarguable importance and interest, dehydrogenation and superhydrogenation atomistic mechanisms are still far from a complete understanding. It is known that dehydrogenation reactions of PAHs are an unfavourable process as it is necessary to overcome an energy barrier (activation energy) which is given by the strength of the R-H bond, usually in the range of 3 – 5 eV for the most typical elements present in organic molecules such as C, N, or O.[Dean1999] These high energy barriers prevent the spontaneous dehydrogenation of PAHs in gas phase at room temperature (RT) *via* R-H bond cleavage.  To overcome this problem, it is a common practice to include catalysts of different nature.[Horváth] However, the barrier, although significantly reduced by the effect of the cata-lyst, can still be high enough to prevent the reaction to proceed at RT. For this purpose, different ways to supply energy such as heat, UV radiation, or variation of the pH have been explored.[review?]

Recently, *On-surface Synthesis* has emerged aiming at describing new covalent reactions yielding unprecedented low-dimensional nanostructures, such as molecules, macromolecules, and low dimensional nanoarchitectures, otherwise unavailable *via* traditional solution-based chemistry.[Gourdon2008] This methodology exploits the ultimate control provided by ultra-high vacuum (UHV) environments on well-defined metal surfaces acting as catalysts.[Held2017][Franc2011][Fan2015][Dong2015] Among the wide collection of available on-surface reactions,[Shen2017][Bjork2014] dehydrogenation and cyclodehydrogenation are among the pioneering and more extensively utilized ones. They have been used in the fabrication of 0D fullerenes,[Otero2008][Amsharov2010] nanodomes, [Sanchez2013][Rim2007] [Sanchez-Valencia2014] and nanographenes; [Simonov2015] [Sanchez-Sanchez2016] [treier2011] [pinardi2013] 1D graphene nanoribbons;[Talirz2016] or even 2D networks and graphene,[Sanchez2015] [Batzill2012] to name some of them. However, this rapidly increasing field suffers from a limited number of substrates where the reaction may occur and, often, the interaction of the reaction product with the metal substrate precludes their further use.

Herein, we report on an unprecedented mechanism for the dehydrogenation and subsequent covalent coupling of organic molecules, namely PAHs, directly on surfaces. It is based on a two-steps process. Firstly, the PAH undergoes superhydrogenation by atomic-hydrogen exposure, thus inducing a weakening of the C-H bond as a consequence of the transition from a $sp^2$ to a $sp^3$ configuration. Secondly, the two hydrogen atoms attached to the same superhydrogenated carbon atom are simultaneously cleaved when facing the corresponding hydrogen atoms of a neighbouring superhydrogenated molecule in a one-to-one reaction that we name "*head-to-head concerted superdehydrogenation (CSDH)*" (see Figure 1). This reaction leads to the formation of new covalent C-C bonds and the release of two hydrogen molecules. Specifically, we show that when surface-adsorbed PAHs –like pentacene or perylene– are annealed in the presence of atomic hydrogen, they are activated and form new covalent macromolecular nanostructures. These processes are characterized by means of low-temperature scanning tunneling microscopy (LT-STM) and rationalized with the help of density functional theory calculations (DFT). The relevance of this work is two-fold. (1) It aims at unveiling the atomistic mechanisms for superhydrogenation and concerted superdehydrogenation reactions of PAHs on surfaces; and (2) it makes use of these ideas to achieve organic nanostructures by PAHs intermolecular covalent coupling. Although there are other examples of H-induced molecular activation,[Son1996] they imply the rupture of C-C bonds within a small cyclic molecule. On the opposite, our method paves the way toward the activation and further covalent coupling of aromatic organic molecules on substrates presenting low reactivity while preserving their internal structure, thus increasing their potential technological interest.

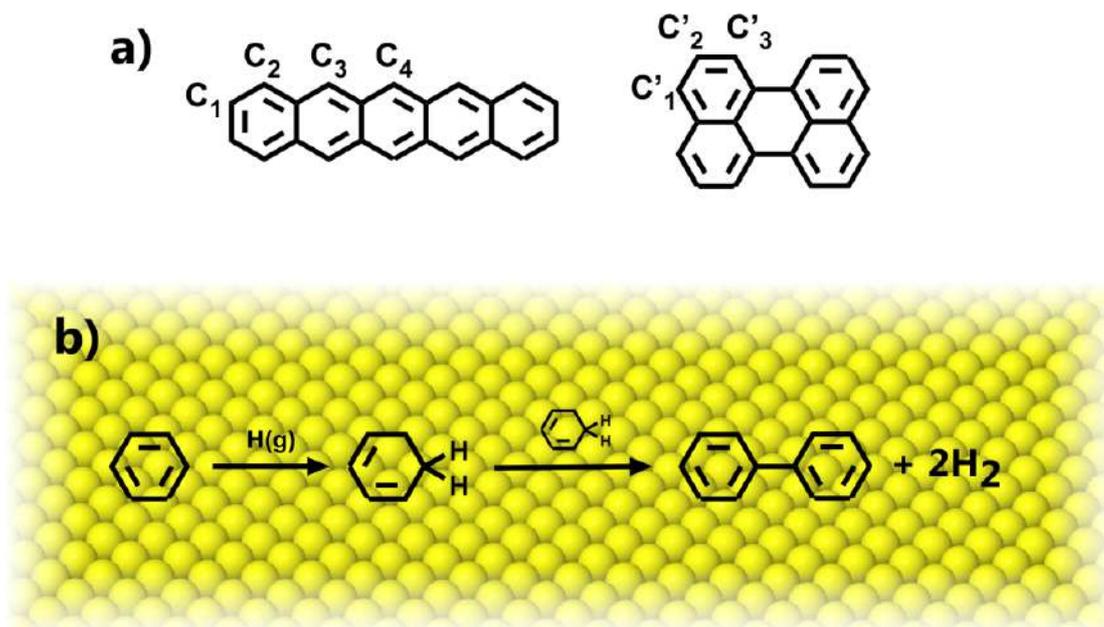

**Figure 1.- a)** Schematic representation of the pentacene (left) and perylene (right) molecules used in these experiments and their inequivalent peripheral C atoms. **b)** Schematic representation of the proposed mechanism resulting in the formation of covalent macromolecular structures and the release of molecular hydrogen.

**Results and Discussion**

We have chosen as model system two prototypical PAHs –pentacene and perylene– and the most commonly used metallic surface in on-surface synthesis –Au(111). The choice of Au is rationalized by the weak molecule-surface interaction, which yields intact molecular adsorption and desorption of the abovementioned PAHs in the temperature range between 375 – 525 K.[France2003] For the sake of clarity, the detailed analysis has been performed with pentacene, while results have been further corroborated with perylene (see SI). Figure 2a shows an STM image of the Au(111) surface after deposition of 1 ML of pentacene molecules at RT. They self-assemble into well-ordered islands with two different domains, closely related to the orientation of the herringbone reconstruction, as previously reported.[France2002] Therefore, the starting point for the experiments hereafter reported will be that kind of 1 ML pentacene-covered Au(111) surface. Post-annealing this system at 525 K under UHV conditions induces the desorption of the vast majority of the molecules (see Figure 2b), in good agreement with reported thermal programmed desorption (TPD) experiments.[France2003] Only some monomers and small clusters remain on the surface, mainly interacting with the elbows of the herringbone reconstruction, the latter being known for their higher reactivity.[ref] Additionally, the inverted appearance of the reconstruction is characteristic of tip-induced motion of isolated molecules along the FCC regions of the surface reconstruction.[Böhringer2000]

However, this scenario changes completely when the annealing of the pentacene layer is carried out in the presence of atomic hydrogen. Figure 2d shows the Au surface after annealing the pentacene layer at 525 K under a continuous supply of atomic hydrogen, followed by a post-annealing at 575 K without further H. A clear increase in the amount of pentacene remaining on the Au(111) surface is observed, in marked contrast to the cases in which no hydrogen or molecular hydrogen are used (Figures 2b and 2c, respectively). This result contrasts with previous efforts for pentacene activation on Au surfaces where, for example, low reaction yields were obtained when using a more reactive surface – Au(110)– and a higher temperature (625 K).[Cui2016] Furthermore, the amount of remaining penta-cene on the surface strongly depends on the atomic hydrogen supplied to the system, as evidenced in Figure 2e, where different hydrogen partial pressures have been used during the annealing. Interestingly, the reaction yield increases with the amount of hydrogen suggesting that, under these experimental conditions, the amount of available H is the reaction limiting factor. A close-view to the so-formed molecular nanostructures shows that they are composed of pentacene molecules arranged into covalent dendritic structures, which are homogeneously distributed around the surface. This assumption is supported by different experimental evidences. Firstly, they remain intact after a 15 nm displacement accompanied by an almost 90º rotation induced by atomic manipulation with the STM tip (red circle in Figure S1). Secondly, so-formed structures present a homogeneous appearance, in good agreement with superimposed scaled covalent schematic models (Figure 3).

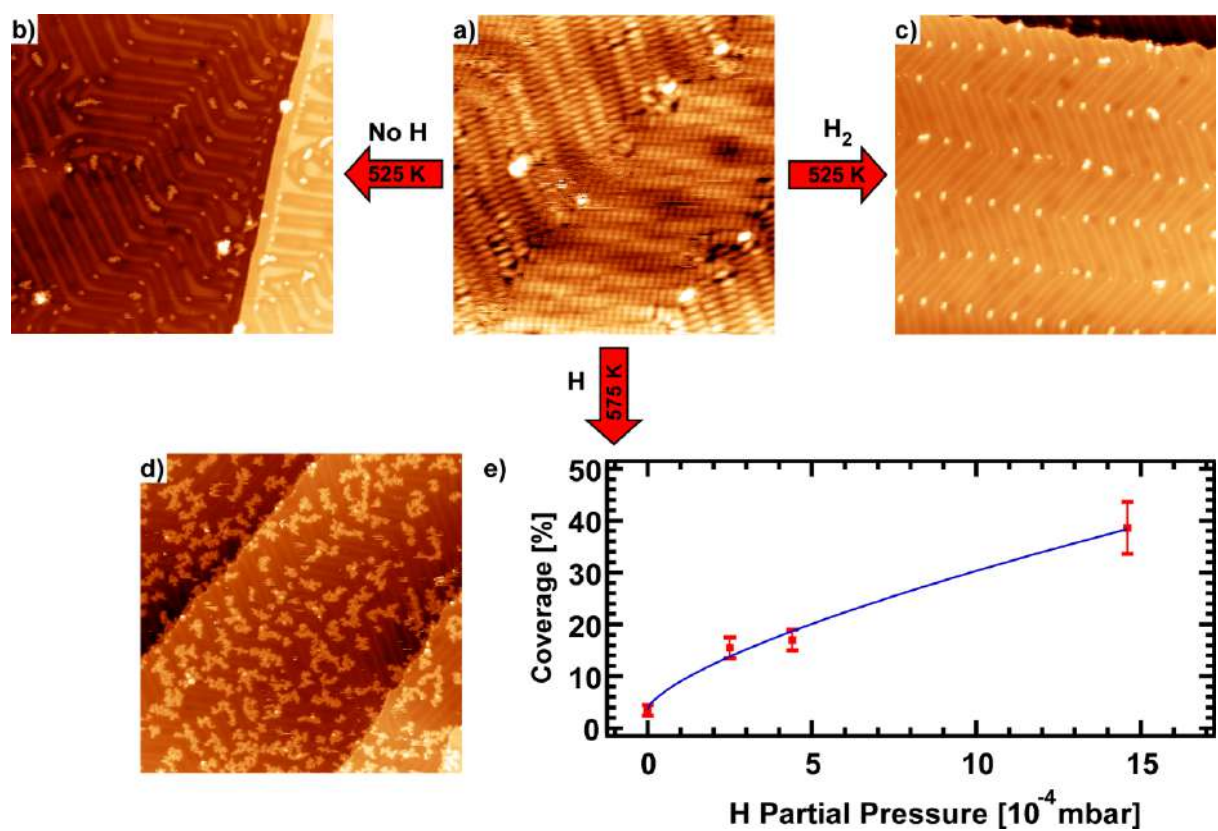

**Figure 2.-** Evolution of 1 ML pentacene on Au(111) with surface temperature and hydrogen partial pressure. **a)** STM image of the as-deposited molecular layer. Individual elongated features correspond to pentacene molecules which self-assemble into two domains following the herringbone reconstruction. STM parameters: 30 nm x 30 nm, I = 20 pA, V = -1.4 V, 77 K. **b)** STM image of the pentacene layer after annealing at 525 K in UHV. Molecules desorb from the Au surface. STM parameters: 100 nm x 100 nm, I = 20 pA, V = -1.5 V, 77 K. **c)** STM image of the pentacene layer after annealing at 525 K in $H_2$ atmosphere (P($H_2$)=5.0 $10^{-7}$ mbar). Molecules desorb from the Au surface. STM parameters: 100 nm x 100 nm, I = 20 pA, V = -1.5 V. **d)** STM image of the pentacene layer after annealing at 525 K in the presence of atomic hydrogen (P(H)=1.5 $10^{-3}$ mbar at the exit of the cracker capillary) and post-annealing at 575 K. Pentacene-related nanostructures appear on the surface as a consequence of intermolecular covalent coupling. STM parameters: 100 nm x 100 nm, I = 20 pA, V = -1.5 V. **e)** Graph showing the evolution of the coverage of pentacene-related nanostructures with hydrogen partial pressure estimated at the exit of the cracker capillary tube (cracker-sample separation ~ 10 cm). The blue line corresponds to a power fit of the experimental data (exponent=0.697). The value found at 0 partial pressure is due to reactions taking place at the elbows of the herringbone reconstruction.

Figure 3a-f shows a set of high-resolution STM images of some of the most characteristic and interesting molecular entities resulting from pentacene annealing under an H flux: L-shape dimers (blue arrows), boomerang-shape dimers (pink arrows), tetramers, and rectangular structures (green and red arrows). The L-shape dimer is proposed to be the result of the covalent coupling at the $C_1$ positions (see Figure 1a) of two pentacene molecules. The assignment of the bonding positions is based on the good agreement between experimental STM image and the schematic scaled model (Figures 3b and 3d). Given the flexibility of the individual C-C bond linking both molecules, the intermolecular angle can adopt different values depending on the surrounding structures and the registry with the surface. As an example, the L-shape dimer in Figure 3a presents an angle of ~90º while the one in panel c is ~120º. On the other hand, the boomerang–shape dimer may correspond to the covalent coupling between $C_1$ and $C_2$ positions of two pentacene molecules (Figures 3d and 3f). In this case, the angle between molecules is ~150º. The tetramer is compatible with the covalent coupling of an L-shape and boomerang-shape dimers as shown in Figure 3b. Finally, we have the rectangular structures, which are probably the most interesting ones (Figures 3e and 3f). These rectangular entities are assigned to nanographene patches formed upon in-registry side-by-side intermolecular coalescence of two (red arrow) or three (green arrows) pentacene molecules. Given their 0D character and their structural atomic precision, these nanographenes are expected to present intriguing electronic and magnetic properties like giant edge state splitting localized at the zigzag edges.[shiyong2016] Interestingly, the yield of nanographene formation is quite high, resulting in 1 nanographene per 60 nm$^2$; in other words, it is possible to find around 60 nanographenes in an area of 60 nm x 60 nm. There exist other examples of on-surface synthesis of atomically precise nanographenes, like those obtained by the surface-assisted intramolecular cyclodehydrogenation of diverse PAHs on metallic surfaces.[Simonov2015][Sanchez-Sanchez2016][treier2011][pinardi2013] There are also several examples of on-surface coalescence of graphene nanoribbons (GNRs) yielding wider, atomically precise GNRs on Au(111), which usually require annealing temperatures above 650 K.[Basagni2015][Deniz2017] However, to the best of our knowledge, these structures are the first realization of nanographenes by on-surface coalescence of unfunctionalized PAHs on a low reactive substrate as Au(111). Furthermore, our hydrogen-assisted strategy decreases the intermolecular merging temperature by around 125 K; that is, a 20 % with respect to other available protocols.

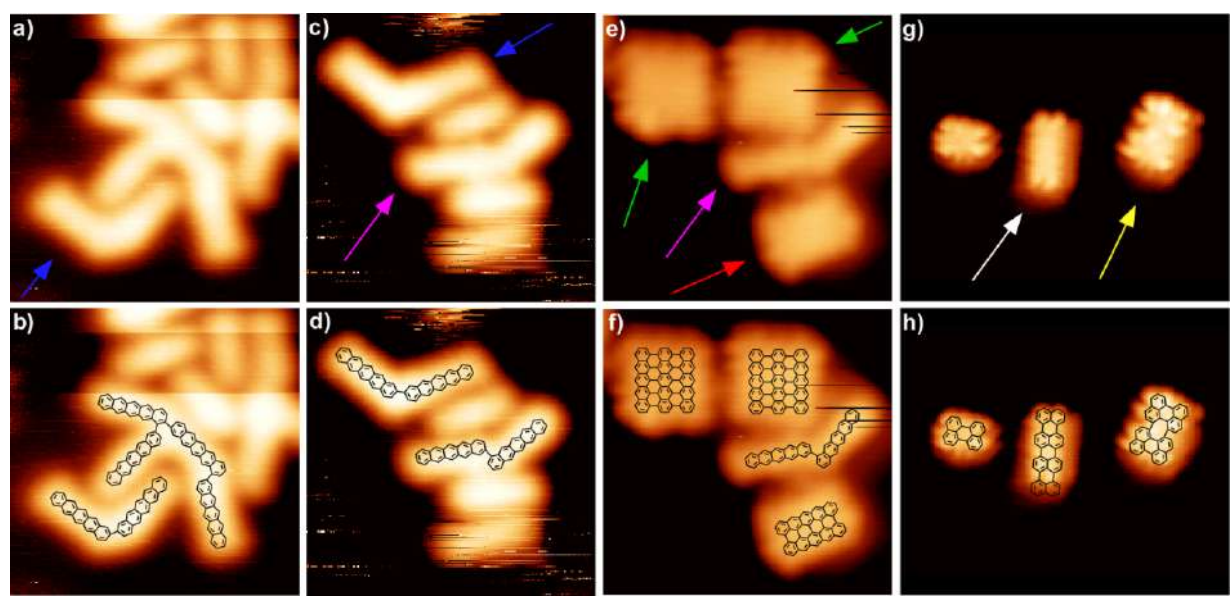

**Figure 3.-** STM images and schematic models for different pentacene- and perylene-related nanostructures formed upon annealing in the presence of atomic hydrogen. **a,b)** L-shape pentacene dimer (blue arrow) and pen-

tacene tetramer formed upon C-C covalent coupling at $C_1$ and $C_2$ positions. STM parameters: 5.0 nm x 5.0 nm, I = 30 pA, V = -1.5 V, 77 K. **c,d)** L- and boomerang-shape (blue and pink arrows, respectively) pentacene dimers formed after activation of $C_1$-$C_1$ and $C_1$-$C_2$ positions, respectively. STM parameters: 5.0 nm x 5.0 nm, I = 20 pA, V = -1.5 V, 77 K. **e,f)** Boomerang-shape pentacene dimer (pink arrow) and rectangular patches (red and green arrows), the latter appearing upon in-registry lateral coalescence of two and three pentacene molecules, respectively. STM parameters: 5.0 nm x 5.0 nm, I = 20 pA, V = -1.5 V, 77 K. **g,h)** Perylene monomer (left) and dimers. Left dimer (white arrow) corresponds to the head-to-head covalent coupling of $C'_1$ carbon atoms while right dimer (yellow arrow) is associated to the lateral fusion of two perylene molecules involving positions $C'_2$ and $C'_3$. STM parameters: 6.0 nm x 7.8 nm, I = 20 pA, V = -1.2 V, 5 K.

To check the universality of this new dehydrogenation methodology, we have carried out experiments using perylene on Au(111). Perylene has been chosen because it combines armchair- and zigzag-edge topologies in similar proportion (see Figure 1a). It has been suggested that zigzag edges in graphene-based structures are more reactive than armchair ones due to the presence of electronic states close to the Fermi level,[Jiang2007] fact that could induce selectivity in the H-mediated dehydrogenation of the molecule. Thus, perylene is an excellent prototype to test this point. As evidenced in Figure S2, our mechanism also works with perylene, resulting in the formation of perylene macromolecules as a result of the H-induced intermolecular covalent coupling. Figure 3g shows an STM image of some interesting and characteristic nanostructures formed when using perylene, namely a monomer and two dimers. High-resolution imaging of the monomer allows unambiguous determination of the molecular orientation; the two maxima at the edges corresponding to the zigzag edge. Thus, we can univocally assign the two dimers to an edge-to-edge fusion of two monomers, i.e., the formation of nanographenes; the one in the center (white arrow) resulting from coalescence along the zigzag edge and the one on the right (yellow arrow), along the armchair edge. This latter case is especially interesting as it would imply the formation of an octagonal ring and two square rings, similar to the structures reported by M. Liu *et al*.[Liu2017] but, once again, without any need for halogen functionalization.

Heretofore, we have proven the possibility to induce PAH dehydrogenation and covalent coupling by on-surface annealing in the presence of atomic hydrogen, as well as that the yield, at 525 K, scales with the amount of available H. However, little is still known about the energetics of the process such as the magnitude of the dehydrogenation barrier necessary to activate the molecules. To get some insight into this point, we have performed experiments in which the annealing temperature has been decreased to 425 K. As it can be observed in the STM images in Figure S3a and b, despite the significant descent of the annealing temperature, it is still possible to find dendritic structures resulting from pentacene dehydrogenation and subsequent intermolecular covalent coupling (blue arrows in Figure S3a and b). Additionally, not all unreacted pentacene molecules have desorbed at this temperature and some of them diffuse over the surface –horizontal strikes in STM images. If the sample is post-annealed at 525 K without further H exposure, unreacted pentacene molecules desorb, making easier the visualization of the so-formed nanostructures (Figure S3c and d). Interestingly enough, the efficiency of the process is reduced at the lower temperature by around 40%, thus indicating the existence of an energy barrier that we estimate directly from Boltzmann probability to be about 0.15 to 0.17 eV. Below, we utilize DFT to assign it to the molecular superhydrogenation barrier.

In order to rationalize our experiments, we have performed extensive DFT calculations to determine the plausibility of the different competing processes involved. Here, we briefly explain the simpler and more direct one (*head-to-head* mechanism), and we defer to the SI the details for a second plausible pathway that we deem as less likely. For the sake of clarity, we have mostly focused our calculations on the pentacene/Au(111) system, performing additional tests for the most relevant involved steps in the napthtalene/Au(111) and perylene/Au(111) interfaces. However, as it will be shown, the proposed mechanisms are general and can be applied to a wide range of PAHs adsorbed on diverse metallic surfaces, although the detailed numbers for each step would vary within the different systems.

Attending to the experimental observations presented in this work, it is clear that the proposed mechanism has to fulfil the following requirements: the reaction needs the presence of atomic hydrogen to proceed; it requires of a surface to adsorb and confine the molecules involved in the reaction – pentacene and H–, thus increasing the encounter probability in comparison to the gas phase, as well as to catalyze the reaction –this rationalizes why such a dehydrogenation process has not been already observed in the gas phase[Montillaud2013]; and it must be kinetically stimulated by temperature. The proposed mechanism, as we will show, consists on the superhydrogenation of two nearby surface-adsorbed molecules and their subsequent concerted superdehydrogenation upon head-to-head collision to end up in the formation of a new C-C bond and two hydrogen molecules.

If we emulate the experiments, in a first step a hydrogen atom approaches a pentacene molecule lying on the surface. When the H atom is close enough, three possibilities arise: (1) the H atom interacts with one of the peripheral H atoms of the molecule, cleaves the C-H, and forms a $H_2$ molecule and a carbon radical, either in a Eley-Rideal or in a Langmuir-Hinshelwood mechanism (see, for example, reference [samorjai1994]); (2) the H atom interacts with one of the peripheral C atoms of the molecule and end ups being adsorbed, thus superhydrogenating the molecule and inducing a local transition from a $sp^2$ to a $sp^3$ configuration; and (3) the H atom interacts with the molecule and it is repelled, leaving it in its canonical form. In case (1), the computed energy barrier is in the range of 0.64-0.74 eV, the lower value corresponding to the central C atom (position $C_4$). This site-dependence energy barrier is in good agreement with the well-known higher reactivity of acenes at their more aromatic central rings.[Schleyer2001] On the other hand, case (2) presents an energy barrier of 0.24-0.26 eV. These two competing processes will coexist; however, given the much lower energy barrier found for case (2), the latter will prevail. Furthermore, molecular superhydrogenation will also favour the subsequent molecular activation necessary to induce intermolecular covalent coupling, as it will work to soften the C-H bond to be broken ($C_{22}H_{14}$+H → $C_{22}H_{15}$; ΔH=-1.5 eV). Such a process is exothermic and transforms the $sp^2$-like bond into a $sp^3$-like one; the two hydrogen atoms attached to the same carbon making approximately a 125.5º angle (in a plane perpendicular to the one of the aromatic ring) and increasing their bonding length to carbon by approximately 2%. The energy stored in the concerned C-H bond decreases approximately from 5.0 eV ($C_{22}H_{14}$ → $C_{22}H_{13}^*$+H ; ΔH=+5 eV) to 3.25 eV per C-H bond ($C_{22}H_{15}$ → $C_{22}H_{13}^*$+2H ; ΔH=+6.5 eV). This is sufficient to make exothermic the dehydrogenation of the PAH when forming two hydrogen molecules and a radical ($C_{22}H_{15}$+2H → $C_{22}H_{13}^*$+$2H_2$ ; ΔH=-2.5 eV). It is obvious that if none of the first two processes take place, then case (3) will occur.

Thus, it seems clear that the initial step involves the superhydrogenation of adsorbed molecules (e.g., $C_{22}H_{14}$+H → $C_{22}H_{15}$). As already mentioned, our DFT on-surface calculations (details in SI) predict a barrier of 0.24-0.26 eV and an enthalpy gain of -1.84 eV, per molecule. These numbers agree with our experimental observation of an initial barrier of approximately 0.17 eV. At this point it is important to remark that the barriers computed within our theoretical framework are always upper bounds. The next step in this head-to-head mechanism consists on the collision at the superhydrogenated positions of two molecules guided by the surface to end up in the formation of a new intermolecular C-C bond and two hydrogen molecules (Fig.4, middle upper panel). This step requires overcoming a barrier of 1.66-1.74 eV (per molecule, depending on the involved C atoms), with an enthalpy gain of 6.25 eV (per C-C bond), e.g. $2C_{22}H_{15}$ → $C_{44}H_{26}$ + $2H_2$. Given that barrier and an attempt frequency of $10^{13}$ Hz (collision rate), we estimate that about half an hour is necessary to complete the process. Finally, the shape of the resulting structure will be determined by the relative position of the involved superhydrogenated molecules, thus explaining the different nanostructures observed in the experiments as no important differences in the energy barriers are obtained.

At this point it should be noted that, due to the particularity of this mechanism where two superhydrogenated species have to bring their superhydrogenated C atoms close enough as to establish a new C-C bond and form two $H_2$ molecules, steric repulsion arguments between hydrogen atoms in a $sp^2$ configuration can decrease the efficiency of the process at central molecular positions like $C_3$ and $C_4$. This effect has been observed for other acenes activated at central positions where no intermolecular C-C bond can be formed and just organometallic complexes, which present longer bonding distances, are achieved.[Urgel2017]

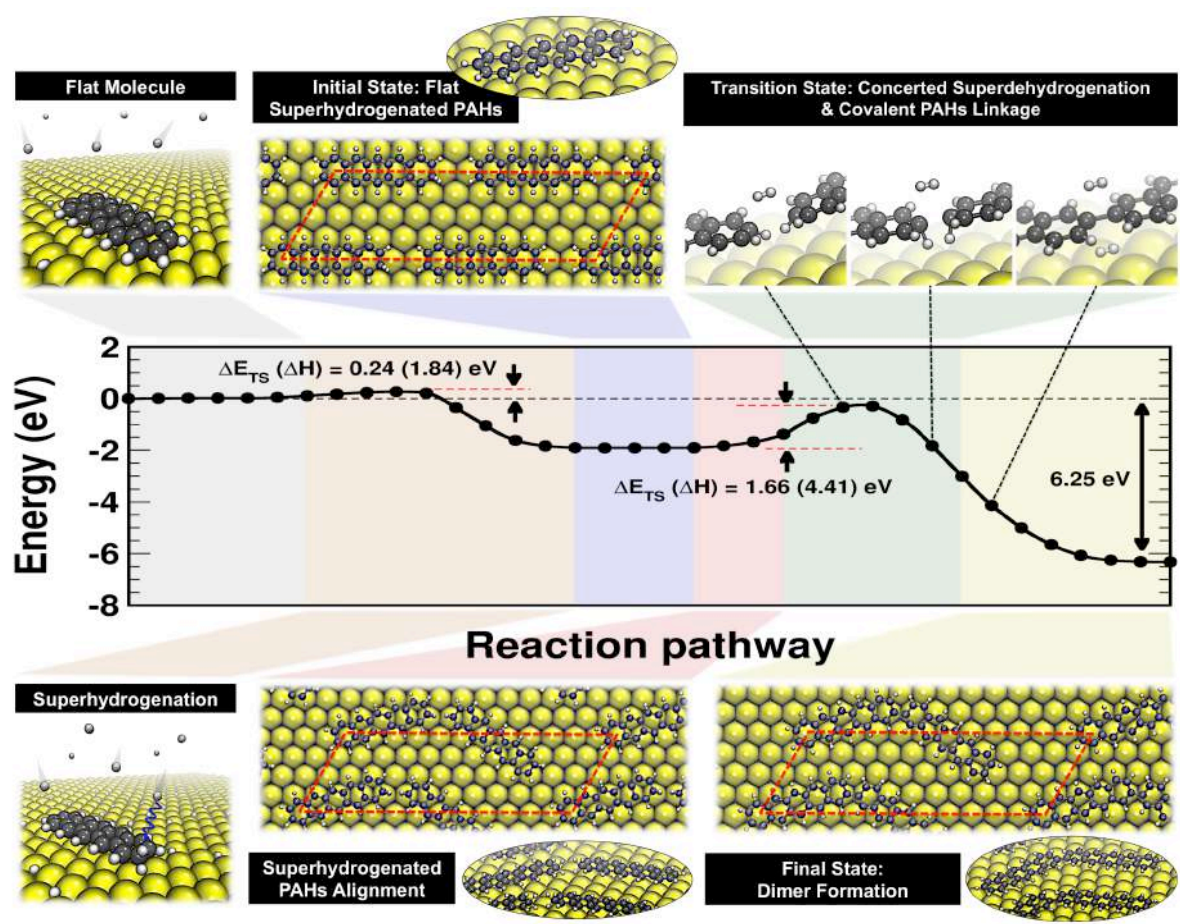

**Figure 4.-** Dimer formation involving $C_1$ positions *via* H-induced superhydrogenation and intermolecular concerted superdehydrogenation of pentacene molecules on Au(111) from CI-NEB calculations (*head-to-head mechanism*). **Top and bottom:** schematic representation of the most important steps involved in the reaction pathway. Yellow, dark grey, and white atoms correspond to Au, C, and H, respectively. Middle: energy diagram associated to the reaction pathway. Shading colors refer to the different steps of the reaction. Red dashed-line boxes superimposed in top views represent the simulation cell used in the calculations.

As already mentioned, DFT calculations indicate that, whichever is the mechanism operating, molecular superhydrogenation is a key requirement in the process. In order to show that molecular superhydrogenation indeed occurs, we have performed an experiment in which molecules have been exposed to atomic hydrogen at RT. Figures S2a and S4 show a set of STM images acquired after exposure of pentacene and perylene to H at RT. As marked by green arrows in Figure S4, it is possible to discern bright features on some of the molecules which are assigned to superhydrogenated species. This result is in good agreement with similar superhydrogenation experiments performed with coronene,[ref] as well as with the presence of $H_2$ species at the edges of graphene nanoribbons (see, for example, Figure S2 in [Ruffieux2016]), thus confirming the existence of superhydrogenated species at RT.

Once we have proved that superhydrogenation takes place, the next step to test our model is to check whether it is possible to activate the molecules by post-annealing a sample where molecules have been superhydrogenated at RT, without further supply of atomic hydrogen. If a sample of superhydrogenated pentacene is post-annealed at 525 K in the absence H, macromolecular nanostructures are formed, in a similar way as those observed when annealing in a flux of atomic H (Fig. S5). If STM images in Figures 2 and S5 are compared, it becomes evident that the final coverage after the whole process is not so different. This is a clear indication that the critical point for the reaction to proceed is the presence of superhydrogenated species as considered in our head-to-head CSDH mechanism.

**Conclusions**

Summarizing, we report on the discovery and rationalization of a novel atomistic mechanism leading to the activation and subsequent covalent coupling of organic molecules on surfaces under UHV conditions to form macromolecular superstructures. It is based on the adsorption of at least an extra H atom at one of their peripheral C atoms (superhydrogenation), thus inducing an $sp^3$ configuration that weakens the C-H bonds. Subsequently, several mechanisms for the activation of the PAHs could operate. In particular, we consider that the most plausible one is the head-to-head CSDH where two superhydrogenated molecules face each other at the superhydrogenation positions, inducing the simultaneous rupture of 4 C-H bonds (CSDH), the formation of 2 $H_2$ molecules, and the creation of a new C-C covalent bond. The so-formed species continue aggregating yielding characteristic dendritic nanostructures. Within the variety of so-formed nanostructures, a non-negligible amount of atomically precise nanographenes is obtained upon side-by-side coalescence of several pentacene molecules.

The proposed mechanism could be responsible for the formation of large aromatic species in circumstellar environments as well as opens a door to a new way of inducing dehydrogenation and covalent coupling of organic molecules on low reactive surfaces, otherwise unavailable due to molecular desorption prior to reaction. We can envision that this method could be exploited to obtain new nanostructures directly on more technologically substrates like semiconductors or insulators. Moreover, we especulate that the use of heteroatomic molecular precursors may lead to new molecular nanoarchitectures with specific and tailored electronic properties.

**Methods**

Experiments have been carried out in a UHV chamber with a base pressure of $1.0 \ 10^{-10}$ mbar, equipped with an LT-STM (Scienta Omicron) and a thermal gas cracker (MGC Series, Mantis Deposition). Au(111) single crystal (Surface Preparation Laboratory, Netherlands) was cleaned using the standard protocol based on repeated cycles of sputtering ($Ar^+$ ions, 1.0 kV) and annealing (~750 K) until judged clean by STM. Pentacene (Sigma Aldrich, 99+%) and perylene (Acros Organics, 98%) molecules were thermally sublimed from a Kentax three-fold evaporator (Kentax GmbH) at a rate of ~1 Åmin$^{-1}$ (sublimation temperatures of 460 K and 400 K, respectively). Prior to deposition, they were thoroughly outgassed for several hours at temperatures 20 K lower than those used for deposition. STM images were acquired in the constant current mode. For the sample preparation, the following procedure was used: a full molecular monolayer of pentacene or perylene was deposited on the clean Au(111) surface held at RT. Then the sample was faced toward the $H_2$ cracker and exposed to a flux of atomic H while ramping the sample temperature from RT to 525K (10 minutes) and then kept it at 525K for another 20 minutes. After that, the annealing was stopped and the sample was allowed to cool down to RT while keeping the atomic H flux during the first 5 minutes. The partial pressure of $H_2$ in the chamber during H exposure was 5.0 10$^{-7}$ mbar unless otherwise stated and the working power of the hydrogen cracker 40 W. Using these parameters, we estimate a cracking efficiency of around 40-50% and a pressure in the capillary around 4 orders of magnitude higher than that in the chamber.

**Acknowledgements**

This work was supported by the Spanish MINECO (Grant MAT2017-85089-C2-1-R), and by the EU via the ERC-Synergy Program (Grant No. ERC-2013-SYG-610256 NANOCOSMOS) and the Innovation Program (Grant 696656: Graphene Core1. Graphene-based disruptive technologies). JIM also acknowledges support from Ramón y Cajal Program (Grant RYC-2015-17730), as well as computing resources from CTI-CSIC.

**References**

[Blumer1976] M. Blumer, "Polycyclic Aromatic Compounds in Nature" *Scientific American* **234**, 34 - 45 (1976)

[Tielens2008] AGGM Tielens, "Interstellar Polycyclic Aromatic Hydrocarbon Molecules" *Annu. Rev. Astron. Astrophys.* **46,** 289–337 (2008)

[Ehrenfreund2006] P. Ehrenfreund, S. Rasmussen, J. Cleaves, and L. Chen. "Experimentally Tracing the Key Steps in the Origin of Life: The Aromatic World" *Astrobiology*. **6,** 490-520 (2006)

[Haritash2009] A.K. Haritash, C.P. Kaushik. "Biodegradation aspects of Polycyclic Aromatic Hydrocarbons (PAHs): A review" *Journal of Hazardous Materials* **169**, 1–15 (2009)

[Habart2004] E. Habart, F. Boulanger, L. Verstraete, C. M. Walmsley and G. Pineau des Forêts. „Some empirical estimates of the H2 formation rate in photon-dominated regions" *A&A* **414**, 531-544 (2004)

[Montillaud2013] J. Montillaud, C. Joblin, and D. Toublanc. "Evolution of polycyclic aromatic hydrocarbons in photodissociation regions" *A&A* **552**, A15 (2013)

[Samorjai1994] G. A. Somorjai, Introduction to Surface Chemistry and Catalysis (Wiley, New York, 1994).

[Schleyer2001] P. R. Schleyer, M. Manoharan, H. Jiao, F. Stahi, Org. Lett. 2001, 3, 3643–3646.

[Merino2014] P. Merino, M.Svec, J.I. Martinez, P. Jelinek, P. Lacovig, M. Dalmiglio, S. Lizzit, P. Soukiassian, J. Cernicharo, J.A. Martin-Gago "Graphene etching on SiC grains as a path to interstellar polycyclic aromatic hydrocarbons formation" *Nature Comm*. **5**, 3054 (2014)

[Dean1999] Lange´s Handbook of Chemistry, John A. Dean, McGraw-Hill, INC. (1999)

[Horváth] Encyclopedia of Catalysis, István T. Horváth, Wiley. Online ISBN: 9780471227618. DOI: 10.1002/0471227617

[Gourdon2008] A. Gourdon "On-Surface Covalent Coupling in Ultrahigh Vacuum" *Angew. Chem. Int. Ed.* **47**, 6950 – 6953 (2008).

[Held2017] Philipp Alexander Held, Harald Fuchs, and Armido Studer. "Covalent-Bond Formation via On-Surface Chemistry" Chem. Eur. J. 2017, 23, 5874 – 5892

[Franc2011] G. Franc, A. Gourdon. "Covalent networks through on-surface chemistry in ultra-high vacuum: state-of-the-art and recent developments" Phys. Chem. Chem. Phys., 2011, 13, 14283–14292

[Fan2015] Q. Fan, J. M. Gottfried, J. Zhu. "Surface-Catalyzed C−C Covalent Coupling Strategies toward the Synthesis of Low-Dimensional Carbon-Based Nanostructures" Acc. Chem. Res. 2015, 48, 2484−2494

[Dong2015] L. Dong, P. N. Liu, N. Lin. "Surface-Activated Coupling Reactions Confined on a Surface" Acc. Chem. Res. 2015, 48, 2765−2774

[Shen2017] Qian Shen, Hong-Ying Gao, Harald Fuchs. "Frontiers of on-surface synthesis: From principles to applications" Nano Today 13 (2017) 77–96

[Bjork2014] J. Bjork, F. Hanke. "Towards Design Rules for Covalent Nanostructures on Metal Surfaces" Chem. Eur. J. 2014, 20, 928 – 934

[Otero2008] Gonzalo Otero, Giulio Biddau, Carlos Sánchez-Sánchez, Renaud Caillard, María F. López, Celia Rogero, F. Javier Palomares, Noemí Cabello, Miguel A. Basanta, José Ortega, Javier Méndez, Antonio M. Echavarren, Rubén Pérez, Berta Gómez-Lor & José A. Martín-Gago. "Fullerenes from aromatic precursors by surface-catalysed cyclodehydrogenation" Nature volume 454, pages 865–868 (2008)

[Amsharov2010] Amsharov, K. et al. Towards the isomer-specific synthesis of higher fullerenes and buckybowls by the surface-catalyzed cyclodehydrogenation of aromatic precursors. Angew. Chem. Int. Edn Engl. 49, 9392–9396 (2010).

[Sanchez2013] Carlos Sánchez-Sánchez, José Ignacio Martínez, Valeria Lanzilotto, Giulio Biddau, Berta Gómez-Lor, Rubén Pérez, Luca Floreano, María Francisca López, José Ángel Martín-Gago. "Chemistry and temperature-assisted dehydrogenation of C60H30 molecules on TiO2(110) surfaces" Nanoscale, 2013,5, 11058-11065

[Rim2007] Kwang Taeg Rim, Mohamed Siaj, Shengxiong Xiao, Matthew Myers, Vincent D. Carpentier, Li Liu, Chaochin Su, MichaelL. Steigerwald, Mark S. Hybertsen, Peter H. McBreen, George W. Flynn,* and Colin Nuckolls. „Forming Aromatic Hemispheres on Transition-Metal Surfaces" Angew. Chem. Int. Ed. 2007, 46, 7891 –7895

[Sanchez-Valencia2014] Juan Ramon Sanchez-Valencia, Thomas Dienel, Oliver Gröning, Ivan Shorubalko, Andreas Mueller, Martin Jansen, Konstantin Amsharov, Pascal Ruffieux, Roman Fasel. "Controlled synthesis of single-chirality carbon nanotubes" Nature volume 512, pages 61–64 (2014)

[Treier2011] Matthias Treier, Carlo Antonio Pignedoli, Teodoro Laino, Ralph Rieger, Klaus Müllen, Daniele Passerone & Roman Fasel. „Surface-assisted cyclodehydrogenation provides a synthetic route towards easily processable and chemically tailored nanographenes" Nature Chemistry volume 3, pages 61–67 (2011)

[Simonov2015] Konstantin A. Simonov, Nikolay A. Vinogradov, Alexander S. Vinogradov, Alexander V. Generalov, Elena M. Zagrebina, Gleb I. Svirskiy, Attilio A. Cafolla, Thomas Carpy, John P. Cunniffe, Tetsuya Taketsugu, Andrey Lyalin, Nils Mårtensson, Alexei B. Preobrajenski. "From Graphene Nanoribbons on Cu(111) to Nanographene on Cu(110): Critical Role of Substrate Structure in the Bottom-Up Fabrication Strategy" ACS Nano, 2015, 9 (9), pp 8997–9011

[Sanchez-Sanchez2016] Carlos Sánchez-Sánchez, Thomas Dienel, Okan Deniz, Pascal Ruffieux, Reinhard Berger, Xinliang Feng, Klaus Müllen, Roman Fasel. "Purely Armchair or Partially Chiral: Noncontact Atomic Force Microscopy Characterization of Dibromo-Bianthryl-Based Graphene Nanoribbons Grown on Cu(111)" ACS Nano, 2016, 10 (8), pp 8006–8011

[pinardi2013] Anna Lisa Pinardi, Gonzalo Otero-Irurueta, Irene Palacio, Jose Ignacio Martinez, Carlos Sanchez-Sanchez, Marta Tello, Celia Rogero, Albano Cossaro, Alexei Preobrajenski, Berta Gómez-Lor, Andrej Jancarik, Irena G. Stará, Ivo Starý, M. Francisca Lopez, Javier Méndez, Jose Angel Mar-tin-Gago. "Tailored Formation of N-Doped Nanoarchitectures by Diffusion-Controlled on-Surface (Cyclo)Dehydrogenation of Heteroaromatics" ACS Nano, 2013, 7 (4), pp 3676–3684

[Talirz2016] Leopold Talirz , Pascal Ruffi eux , Roman Fasel. "On-Surface Synthesis of Atomically Precise

Graphene Nanoribbons" Adv. Mater. 2016, 28, 6222–6231

[Sanchez2015] Carlos Sánchez-Sánchez, Sebastian Brüller, Hermann Sachdev, Klaus Müllen, Matthias Krieg, Holger F. Bettinger, Adrien Nicolaï, Vincent Meunier, Leopold Talirz, Roman Fasel, and Pascal Ruffieux. "On-Surface Synthesis of BN-Substituted Heteroaromatic Networks" ACS Nano, 2015, 9 (9), pp 9228–9235.

[Batzill2012] Matthias Batzill. "The surface science of graphene: Metal interfaces, CVD synthesis, nanoribbons, chemical modifications, and defects" Surface Science Reports 67 (2012) 83–115

[Son1996] "Gas Phase Atomic Hydrogen Induced Carbon−Carbon Bond Activation in Cyclopropane on the Ni(100) Surface" Kyung-Ah Son, and John L. Gland. J. Am. Chem. Soc., 1996, 118 (43), pp 10505–10514

[France2003] C. B. France, P. G. Schroeder, J. C. Forsythe, B. A. Parkinson. "Scanning Tunneling Microscopy Study of the Coverage-Dependent Structures of Pentacene on Au(111)" Langmuir 2003, 19, 1274-1281

[France2002] C. B. France, P. G. Schroeder, B. A. Parkinson. "Direct Observation of a Widely Spaced Periodic Row Structure at the Pentacene/Au(111) Interface Using Scanning Tunneling Microscopy" Nano Lett., 2 (2002) 693-696

[Böhringer2000] Matthias Böhringer, Karina Morgenstern, Wolf-Dieter Schneider, Richard Berndt. „Reversed surface corrugation in STM images on Au(111) by field-induced lateral motion of adsorbed molecules" Surface Science 457 (2000) 37–50

[Cui2016] Ping Cui, Qiang Zhang, Hongbin Zhu, Xiaoxia Li, Weiyi Wang, Qunxiang Li, Changgan Zeng, Zhenyu Zhang. "Carbon Tetragons as Definitive Spin Switches in Narrow Zigzag Graphene Nanoribbons" Phys. Rev. Lett. 116, 026802 (2016)

[Shiyong2016] Shiyong Wang, Leopold Talirz, Carlo A. Pignedoli, Xinliang Feng, Klaus Müllen, Roman Fasel, Pascal Ruffieux. "Giant edge state splitting at atomically precise graphene zigzag edges" Nature Communications 7, 11507 (2016)

[Basagni2015] Andrea Basagni, Francesco Sedona, Carlo A. Pignedoli, Mattia Cattelan, Louis Nicolas, Maurizio Casarin, Mauro Sambi. "Molecules–Oligomers–Nanowires–Graphene Nanoribbons: A Bottom-Up Stepwise On-Surface Covalent Synthesis Preserving Long-Range Order" J. Am. Chem. Soc., 2015, 137 (5), pp 1802–1808

[Deniz2017] Okan Deniz, Carlos Sánchez-Sánchez, Tim Dumslaff, Xinliang Feng, Akimitsu Narita, Klaus Müllen, Neerav Kharche, Vincent Meunier, Roman Fasel, Pascal Ruffieux. "Revealing the Electronic Structure of Silicon Intercalated Armchair Graphene Nanoribbons by Scanning Tunneling Spectroscopy" Nano Lett., 2017, 17 (4), pp 2197–2203

[Jiang2007] De-en Jiang, Bobby G. Sumpter, Sheng Dai. "Unique chemical reactivity of a graphene nanoribbon's zigzag edge" The Journal of Chemical Physics 126, 134701 (2007)

[Liu2017] Meizhuang Liu, Mengxi Liu, Limin She, Zeqi Zha, Jinliang Pan, Shichao Li, Tao Li, Yangyong He, Zeying Cai, Jiaobing Wang, Yue Zheng, Xiaohui Qiu, Dingyong Zhong. "Graphene-like nanoribbons periodically embedded with four- and eight-membered rings" NATURE COMMUNICATIONS 8, 14924 (2017)

[Coronene]

[Urgel2017] J. I. Urgel, H. Hayashi, M. Di Giovannantonio, C. A. Pignedoli, S. Mishra, O. Deniz, M. Yamashita, T. Dienel, P. Ruffieux, H. Yamada, R. Fasel. "On-Surface Synthesis of Heptacene Organometallic Complexes" J. Am. Chem. Soc. 139 (34), 11658–11661 (2017). DOI: 10.1021/jacs.7b05192

[Ruffieux2016] "On-surface synthesis of graphene nanoribbons with zigzag edge topology". Pascal Ruffieux, Shiyong Wang, Bo Yang, Carlos Sánchez-Sánchez, Jia Liu, Thomas Dienel, Leopold Talirz, Prashant Shinde, Carlo A. Pignedoli, Daniele Passerone, Tim Dumslaff, Xinliang Feng, Klaus Müllen and Roman Fasel. Nature 531, 489–492 (2016)

40 referencias